\documentclass[aip,pop,reprint,graphicx]{revtex4-1}

\usepackage{graphicx}
\usepackage{xcolor}
\usepackage[normalem]{ulem}

\draft 

\begin{document}

\title{Shocks propagate in a 2D dusty plasma with less attenuation than that due to gas friction alone} 

\author{Anton\ Kananovich}

\affiliation{Department of Physics and Astronomy, University of Iowa, Iowa City, Iowa 52242}

\author{J.\ Goree}
\date{\today}

\begin{abstract}
In a dusty plasma, an impulsively generated shock, \textit{i.e.}, blast wave, was observed to decay less than would be expected due to gas friction alone. In the experiment, a single layer of microparticles was levitated in a radio-frequency glow-discharge plasma. In this layer, the microparticles were self-organized as a 2D solid-like strongly coupled plasma, which was perturbed by the piston-like mechanical movement of a wire. To excite a blast wave, the wire's motion was abruptly stopped, so that the input of mechanical energy ceased at a known time. It was seen that, as it propagated across the layer, the blast wave's amplitude persisted with little decay. This result extends similar findings, in previous experiments with 3D microparticle clouds, to the case of 2D clouds. In our cloud, out-of-plane displacements were observed, lending support to the possibility that an instability, driven by wakes in the ion flow, provides energy that sustains the blast wave's amplitude, despite the presence of gas damping. 

\end{abstract}

\maketitle 

\section{Introduction}
A dusty plasma consists of electrons, ions, neutral gas, and microparticles.~\cite{bellan2008fundamentals,piel2010dusty,fortov2007physics,shukla2001introduction,tsytovich2008elementary,vladimirov2005physics,melzer2019physics,konopka2016guest} Electrons and ions collect on the surface of the microparticles, which accumulate a large charge of thousands of elementary electron charges. Because of that large charge, the microparticles can behave as a strongly coupled plasma component, organizing in a crystalline microstructure.~\cite{chu1994direct,thomas1994plasma,hayashi1994observation,melzer1994experimental,hartmann2010crystallization,gogia2017emergent,kostadinova2018transport,hariprasad2020experimental} Also because of that large charge, large amplitudes are easily attained when compressional waves propagate through a dusty plasma. Wave amplitudes are often great enough for the wave to become nonlinear, and in some cases they can have properties of a shock.~\cite{samsonov1999mach,melzer2000laser,samsonov2003kinetic,samsonov2004shock,fortov2005shock,heinrich2009laboratory,jiang2009mach,nosenko2009dynamics,merlino2012dusty,saitou2012bow,usachev2014externally,jaiswal2016experimental,sharma2016observation,jaiswal2018experimental,ghosh2006large,xue2007non,ghosh2008damped,ghosh2008longitudinal,asgari2009effects,ghosh2009shock,mamun2009dust,mamun2009cylindrical,mamun2009formation,asgari2011dust,das2012formation,asgari2013dust,wang2016nonadiabatic,dev2017complex,charan2018supersonic,li2001shock}

Shocks can in general be classified as continuously driven (\textit{i.e.}, piston-generated) or impulsively driven (\textit{i.e.}, blast waves). In the case of a continuously driven shock wave, there is a constant energy input into the system, so that the amplitude of the propagating shock is sustained and does not decay. In contrast, a blast wave results from an impulsive deposition of energy, which has a finite time duration and occurs within a localized volume. After this initial impulsive energy input, the blast wave propagates without any further input of energy from an external source. In this paper we report an experiment with a blast wave in dusty plasma; we previously reported a different experiment with a continuously driven shock.~\cite{kananovich2020experimental}

A shock's energy generally experiences not only an external input, but also a dissipation. The dissipation mechanism that dominates laboratory dusty plasmas is usually friction exerted on the microparticles by neutral gas.~\cite{nunomura2002dispersion,flanagan2010observation,flanagan2011development,fortov2003dust,thomas2006measurements,trottenberg2006dust,schwabe2007highly,sheridan2008experimental} This gas-friction dissipation mechanism was identified experimentally as a cause of damping not only for blast waves,~\cite{samsonov2004shock} but also other longitudinal and transverse waves~\cite{nunomura2002dispersion} and solitons as well.~\cite{sheridan2008experimental} Because all laboratory dusty plasmas include neutral gas, one might expect a blast wave to display an attenuation, with its amplitude decaying as it propagates, after the initial impulsive energy input.

Unexpectedly, however, a lack of attenuation was observed, despite the dissipative effects of gas friction, in a previous dusty plasma experiment by \citeauthor{fortov2004large},~\cite{fortov2004large} with an impulsively generated large-amplitude pulse. Their cloud of microparticles filled a three-dimensional (3D) volume within a low-temperature DC-discharge plasma. In that experiment, the large-amplitude compressional pulse wave was excited externally by a sudden movement of gas, which was produced by a moving plunger. \citeauthor{fortov2004large} reported that their high-amplitude pulse propagated for 500~ms without significant attenuation, even though a much stronger attenuation, requiring only 15~ms, would be expected due to gas friction, given the high gas pressure used in that experiment.

This unexpected lack of attenuation of their blast wave requires an explanation, as \citeauthor{fortov2004large}~\cite{fortov2004large} noted. Because their high gas density would be expected to cause a ``great damping,'' they remarked that their observation of an unattenuated waveform indicates that ``the wave must have an energy source other than the initial impulse." In their conclusion, they suggested that their wave was driven by an instability, and they suggested the dust-acoustic instability could serve as the mechanism for an energy source. (We note that the dust-acoustic instability's energy source is provided by a flow of ions passing through the cloud of microparticles.~\cite{winske1999numerical,rosenberg2008note,tadsen2015self}) This lack of attenuation, in the experiment of \citeauthor{fortov2004large},~\cite{fortov2004large} also captured the attention of other authors, \citeauthor{fortov2010complex};~\cite{fortov2010complex} they stated that the almost-undamped propagation suggests a mechanism of strong energy influx, and hinted at an alternative mechanism, a modulational instability. \citeauthor{fortov2010complex} also reviewed an earlier experiment by \citeauthor{samsonov2003kinetic},~\cite{samsonov2003kinetic} who similarly used a momentary gas flow to disturb a 3D microparticle cloud, but in an RF-discharge plasma. In that experiment, there might have also been an unexpected lack of attenuation, according to \citeauthor{fortov2010complex}.

In this paper we report another observation of little decay, for a blast wave, despite the dissipative effects of gas friction. Our experiment used a different kind of dusty plasma, one with a single two-dimensional (2D) layer of microparticles. As in the experiment of \citeauthor{fortov2004large}, our electron-ion plasma filled a 3D volume, but unlike their experiment, we prepared our microparticle cloud to be limited to a 2D monolayer, not a 3D volume. Our experiment demonstrates that the previously reported result for 3D microparticle clouds, that a shock's decay is much less than would be expected if gas friction alone acts on the microparticles, occurs also in 2D microparticle clouds.

Experiments with two-dimensional clouds can be different from those with 3D clouds, for several reasons.  Importantly, the possible energy mechanisms for a 2D cloud can be different from those in a 3D cloud. Although a 2D cloud can sustain a compressional shock wave, just as a 3D cloud can, the instabilities are not entirely the same in 2D and 3D clouds. In either case, instabilities can be driven by ion flow, but the shock propagation cannot be parallel to the ion flow in a 2D cloud, which is unlike the situation in a 3D cloud. In a 2D cloud, the ion flow must be nearly perpendicular to the 2D layer, due to the direction of the electric field that serves the dual purpose of levitating the microparticle cloud and driving the ion flow. For a 2D cloud, instabilities driven by ion flow have been well studied experimentally and theoretically, and there is a  theory for the Schweigert instability.~\cite{piel2017plasma,melzer1996experimental,mukhopadhyay2014experimental,melzer1996structure,shveigert2000melting,schweigert1998plasma} This is different from the ion-acoustic instability in a 3D cloud, which involves ion flow that is aligned with the direction that a wave propagates within a 3D volume.

Another difference, for experiments with 2D clouds, is the greater ease of making a crystalline microstructure than in 3D clouds. A stable crystalline microstructure can in general be attained only by using enough gas friction to suppress instabilities, and in a 2D cloud this can be done at a much lower gas friction level than is possible in a 3D cloud. Strong-coupling effects such as a crystalline microstructure can be observed in a 2D cloud~\cite{melzer2019physics,pieper1996experimental,schella2011melting,arp2004dust,sheridan2016self,haralson2016laser} even when the gas pressure is two orders of magnitude less than in a 3D cloud. Moreover, the microstructure can be observed more easily in a 2D cloud, because it is possible to focus a camera so that it views all the microparticles, which is not practical in a 3D cloud using ordinary imaging.

One feature of our blast-wave experiment is that our mechanical energy input stops at a definitive time. This is important so that we can exclude the possibility that a greatly reduced attenuation is somehow due to a lingering of this mechanical energy input, during the propagation. For example, in the case of previous impulsive experiments where the mechanical energy input came from a gas puff, although it was known when the gas valve was closed, it was not necessarily known exactly when the gas has stopped circulating. To provide an impulsive mechanical energy input that ceases at a definitive time, we will use a moving wire as the exciter. The wire's motion is precisely controlled by a stepper motor, and the time that its motion stops is confirmed by our video imaging.

\section{Experiment}

\subsection{Dusty plasma}

\begin{figure}[ht]
\includegraphics[width=0.9\columnwidth]{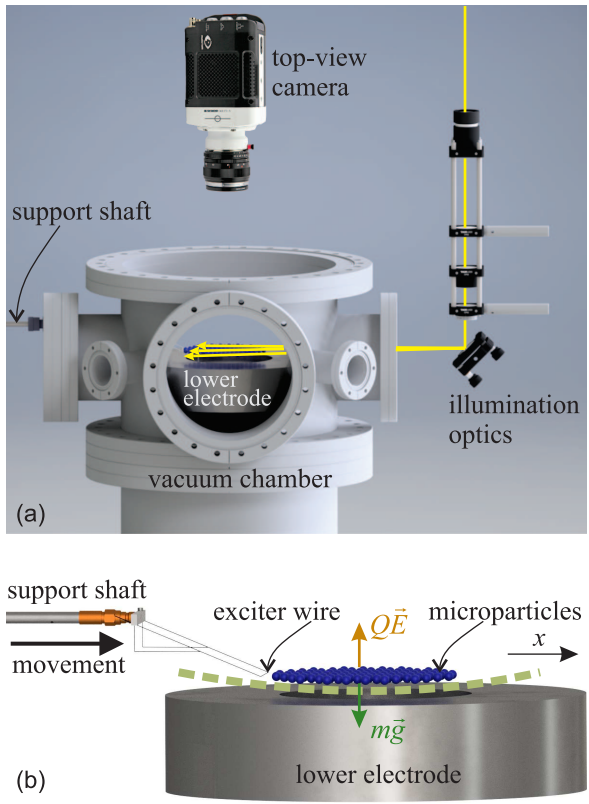}
\caption{\label{fUnexpectedSetup}Experimental setup. (a) Schematic of the vacuum chamber, shown without flanges. The microparticles were imaged simultaneously by a top-view camera with illumination by a horizontal laser sheet, and a side-view camera (not shown) which used separate illumination optics. (b) Schematic of the microparticle cloud and the manipulation setup. A horizontal exciter wire was propelled along the $+x$~direction. The wire and its support structure are shown to scale, while other features are not. The microparticle cloud was levitated in a plasma sheath, which (as sketched by a dashed line) had a curved edge,  conforming to a shallow depression in the lower electrode. The argon plasma, which filled the chamber, was sustained by applying radio-frequency voltage between the lower electrode and the grounded chamber walls. This experimental setup was similar to the one in Ref.~\cite{kananovich2020experimental}, except that the chamber, lower electrode, and the wire's support structure were all different.}%
\end{figure}

A single layer of microparticles was levitated above the lower electrode in our modified GEC chamber, Fig.~\ref{fUnexpectedSetup}. The chamber was configured as in~Ref.~\cite{haralson2017overestimation}, but with the addition of a moving wire, for exciting the blast wave.
Radio-frequency power, at 13.56~MHz, was capacitively coupled to the lower electrode, where the peak-to-peak and self-bias voltages were $-117$ and 149~V, respectively. These voltage measurements remained steady within $\pm 3$~V during the experiment. Argon gas was used at a pressure of 17.0~mTorr, with a small flow rate of 0.15~sccm, chosen to avoid disturbing the microparticles while maintaining gas purity.

The microparticles were melamine-formaldehyde spheres.\footnote{purchased from Microparticles GmbH (Berlin)} By introducing a limited quantity of these microparticles, we were able to prepare a single-layer two-dimensional (2D) cloud, located above the lower electrode. Our microparticles' large diameter, 8.69~$\mu$m, helped them to collect a large charge, which we measured as $Q = -1.5 \times 10^4e$, by analyzing video recordings of microparticle motion in a crystal, as in Ref.~\cite{kananovich2020experimental}. This analysis also yielded other parameters for the microparticle cloud, including the longitudinal sound speed, which was determined to be 16~mm/s. This sound speed will be used for calculating Mach numbers in our shock experiments.

\subsection{Imaging}

Our primary diagnostic tool was video microscopy.~\cite{feng2007accurate,feng2011errors,feng2016particle} Most of our data came from the top-view camera, which was a 12-bit high-speed Phantom Miro~M120. This top-view camera was operated at 70~frames/s for our measurements of the crystal, and a faster rate of 800~frames/s for our runs with blast waves. The spatial resolution was 29.79~pixels/mm. The camera's lens was covered by a bandpass filter, blocking light at wavelengths different from the 577-nm laser light, which was shaped into a sheet to illuminate the microparticles. The sheet's thickness was chosen to be 1.0~mm, which is wider than is typically used in 2D dusty plasma experiments, so that microparticles could not move out of the laser sheet, which was a difficulty in previous shock experiments.~\cite{samsonov2004shock}

To allow the detection of out-of-plane displacements, we also used a side-view camera, unlike the 2D shock experiment of \citeauthor{samsonov2004shock}~\cite{samsonov2004shock} Detecting out-of-plane displacements is significant because they are essential for the Schweigert instability. This instability arises from ion wakes downstream of a microparticle.~\cite{piel2003waves,melzer1996structure,schweigert1996alignment,schweigert1998plasma} These wakes have the greatest effect when a microparticle is located slightly above or below another microparticle, for the case of a vertical ion flow.

Our side-view camera was a Basler Pilot piA~1600~-~35gm, the same as in the PK-4 flight instrument.~\cite{pustylnik2016plasmakristall} To allow operating this camera at 100~frames/s, we recorded images not for the entire sensor, but only for the  $1600 \times 100$ pixel portion of the sensor where the image of the microparticle layer was located. The imaging setup for this side-view camera used a vertical sheet of light, to illuminate the particles. This vertical sheet was produced by a 632.8-nm HeNe laser, and a matching bandpass filter was fitted to the camera. This vertical sheet, which illuminated a cross section of the horizontal layer of microparticles, had a thickness of about 1~mm, which was chosen to be larger than the interparticle spacing. Compared to our main camera, which viewed from the top, this side-view camera had a more limited spatial resolution of only 13.12~pixels/mm.

The top-view and side-view cameras were synchronized using an external clock. Since the top-view camera was operated at a frame rate eight times greater than the side-view camera, the latter was triggered simultaneously with every ninth frame of the top-view camera. The two cameras were aligned and calibrated so that, in their images, the $x$-coordinates correspond. This calibration was done by imaging a test object~(see supplementary material~\cite{supp}), which was placed in the field of view of both cameras.

\subsection{Gas damping}

The gas damping rate will be an important parameter later, when we analyze our experimental results. Here we present a range of values for the theoretical gas damping rate $\nu_E$, for our experimental conditions. The subscript $E$ reflects the Epstein theory of gas damping, which we assume.

We consider the theoretical case where gas friction is the only force acting on a microparticle. The force acting on a particle is
\begin{equation}
F_g=\nu_E m_p v_p,
\label{eqUnexpectedDrag}
\end{equation}
where $m_p$ and $v_p$ are the microparticle's mass and speed, respectively. Calculating the ratio of the force and the microsphere's momentum $m_p v_p$, we obtain the damping rate
\begin{equation}
\nu_E = \delta \sqrt{\frac{8{{m}_{g}}}{\pi {{k}_{B}}{{T}_{g}}}}\frac{p_g}{\rho_p r_p},
\label{eqUnexpectedNu}
\end{equation}
that would be expected if gas friction alone altered the microsphere's energy. Here, $\rho_p$ and $r_p$ are the mass density and radius of the microparticle, while $p_g$, $m_g$ and $T_g$ are respectively the pressure, atomic mass, and temperature of the gas. The leading coefficient in Eq.~(\ref{eqUnexpectedNu}) must have a value in the range $1 \le \delta \le 1.442$, depending on how the gas atoms reflect from a sphere's surface, according to the Epstein theory.~\cite{fuchs1964mechanics,epstein1924resistance}

For our experimental conditions, the gas damping rate in Eq.~(\ref{eqUnexpectedNu}) must be in the range

\begin{equation}
2.21 \le \nu_E \le 3.19 \text{~s}^{-1}.
\label{eqUnexpectedNuBord}
\end{equation}
The values that bracket the range in Eq.~(\ref{eqUnexpectedNuBord}) correspond to the limiting values of 1 and 1.442 for the coefficient $\delta$.

As an experimental confirmation of the theoretical range of Eq.~(\ref{eqUnexpectedNuBord}), we analyzed sloshing-mode data from our experiment. Doing this, we found a gas damping rate of 2.9 s${}^{-1}$, as explained in the supplementary material.~\cite{supp}

\subsection{Blast wave generation}

In order to generate a blast wave, \textit{i.e.}, an impulsively driven shock wave, we require an excitation mechanism that has a finite time duration. For this purpose, in previous dusty plasma experiments a momentary gas flow was used in a 3D cloud,~\cite{samsonov2003kinetic} while an electrical pulse was applied to a stationary wire to generate a blast wave in a 2D microparticle cloud.~\cite{samsonov2004shock} Our approach with a 2D cloud also used a wire, but it was not stationary, but instead it was moved and then abruptly stopped. This approach had the advantage that the energy input stopped at a known time, because we detected the stopping of the wire's motion.

An electrical repulsion, between the wire and the microparticles, allowed us to manipulate the microparticles. The microparticles were floating electrically, so that they had a negative potential as compared to the surrounding plasma. Similarly, our wire was always allowed to float, as we did not apply an external voltage to it. As the wire moved, it acted like a piston since nearby microparticles were repelled from the wire. By abruptly stopping the motion of the wire, analogous to stopping the motion of a piston in a gas cylinder, we were able to excite a blast wave. This method of blast-wave generation differs from that of Ref.~\cite{samsonov2004shock}, where the wire was stationary, and a step-wise change in electrical potential was applied to the wire to repel nearby particles impulsively.  It is also different from the shock-wave excitation mechanism we used in our previous experiment, Ref.~\cite{kananovich2020experimental}, where the wire was moved steadily, analogous to a piston moving at a steady speed in a gas cylinder, so that a shock was driven continuously rather than as a blast wave.

The mechanical structure of the exciter wire was similar to that in our previous experiment, Ref.~\cite{kananovich2020experimental}.  For the present experiment, the wire had the same diameter of 0.41~mm but a greater length of 60~mm. The exciter wire was propelled horizontally by the same motor drive as in Ref.~\cite{kananovich2020experimental}, although the motor drive was programmed differently, so that the wire's motion was stopped abruptly in the present experiment. We also used a different plasma chamber. The chamber, and the wire's structure, are shown in photographs in the supplementary material.~\cite{supp}

The wire's inward horizontal motion was abruptly stopped at a time that was well determined. Importantly, this time was known because the motor drive and cameras were synchronized, and the wire was visible in the cameras' fields of view. As we mentioned earlier, since our wire's motion was abruptly ceased, we would expect the energy input from its mechanical motion to also cease, so that the amplitude of the propagating compressional pulse would thereafter decay due to gas friction, if other energy input mechanisms are absent.

The sequence of experimental steps were as follows. Before the first experimental run, we allowed the microparticle cloud to anneal into a crystalline structure, and during this unperturbed time we  recorded its random thermal motion. Next, we performed a sequence of experimental runs with manipulation. At the beginning of each run, the microparticle cloud had a crystalline structure; the exciter wire was located outside the cloud and it was at rest. We then started the wire's inward horizontal motion, at a steady speed having a desired value. Video recording began, triggered by a photogate that sensed the wire's motion. During this manipulation,  the wire moved toward the microparticle cloud, the cloud was disturbed, and a compressional pulse propagated away from the wire. This propagation started at the cloud's edge, and it continued in the same $+x$ direction as the wire's motion. As the moving wire approached a specified position, approximately where the cloud's outer edge was originally located, we brought the wire to a halt with a rapid acceleration of $-360$~mm/s${}^2$. After the wire's motion was halted, the compressional pulse continued propagating across the cloud, yielding the main data analyzed in this paper. The pulse propagated across the entire cloud, and then the video recording ended. To prepare for the next experimental run, the exciter wire was retracted to its original position, allowing the microparticle cloud to relax to its original location. We waited at least 15~min, allowing the cloud to anneal again into a crystalline microstructure, before the next run. After the sequence of experimental runs was completed, as a final step, the unperturbed random motion of a crystalline structure was recorded again.

\section{Results}

\begin{figure*}[ht]
	\includegraphics[width=0.9\textwidth]{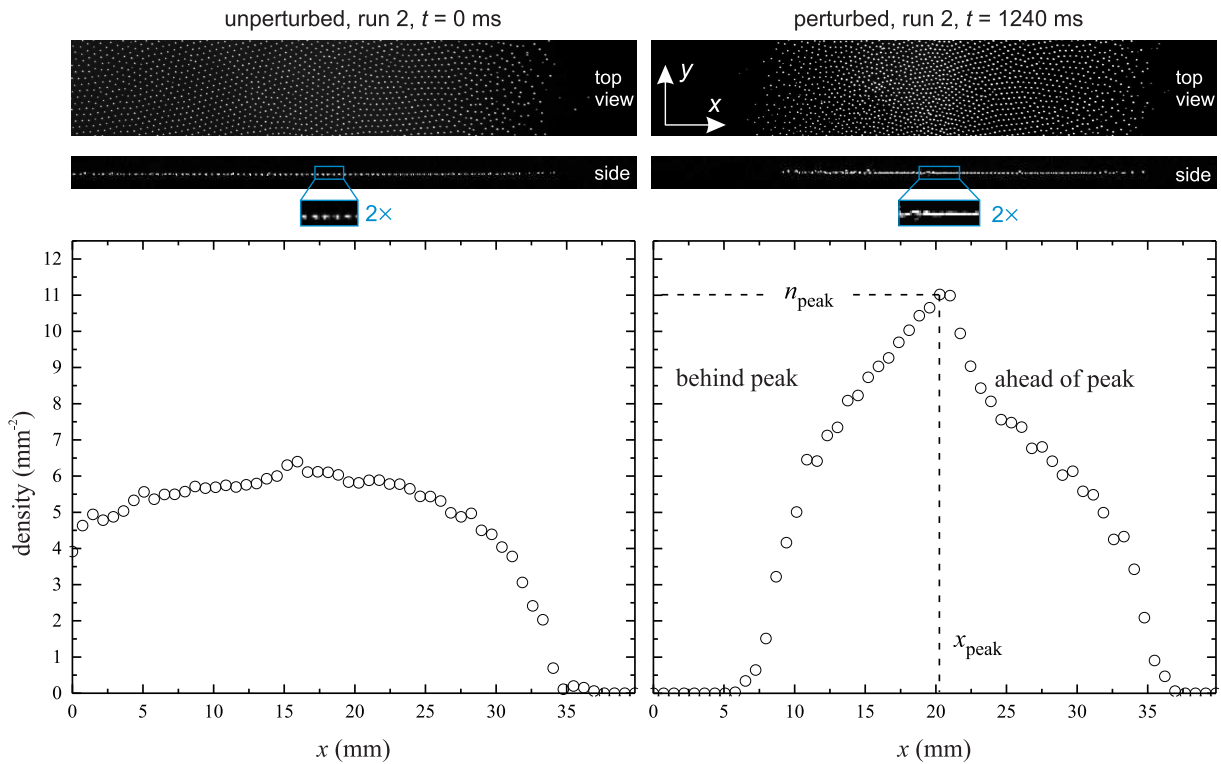}
\caption{\label{fUnexpected2camviews}Images of the microparticle cloud, and corresponding number-density profiles, in the region of interest. Data are shown in two columns; the unperturbed condition is at an early time, before the cloud was disturbed by the exciter wire, while the perturbed condition is at a time after the wire had approached the cloud and then stopped. In the perturbed condition, a compression in the cloud propagated to the right. In this pulse, the number density was compressed nearly two-fold, and its profile had a strong gradient on the peak's leading edge. The side-view camera images confirm that the cloud was a single layer in the unperturbed condition. In the perturbed condition, although it remained as a single layer ahead of the peak, the cloud experienced out-of-plane displacements near the pulse's peak and behind it. These data are for run~2, in which the wire was stopped at $t = 990$~ms after moving at a speed of 50.8~mm/s. Data for all times in this run can be seen in a video in the supplementary material.~\cite{supp}}
\end{figure*}

Figure~\ref{fUnexpected2camviews} shows data for the particle cloud, with images from the top-view and side-view cameras, and the density profile. Images from the top-view camera, shown in the upper panels, reveal the overall arrangement of microparticles. These top-view images are cropped to show the region of interest that we will analyze further. Images from the side-view camera are used to detect any out-of-plane displacements. The bottom panels of Fig.~\ref{fUnexpected2camviews} are profiles of the areal density $n$, obtained from top-view images by counting microparticles located within rectangular bins. The bin width of 0.724~mm corresponds to three Wigner-Seitz radii. In each bin, the number of counts was weighted using a cloud-in-cell algorithm.~\cite{birdsall1985plasma}

\subsection{Comparing unperturbed and perturbed conditions}
In~Figure~\ref{fUnexpected2camviews}, the panels on the left show the microparticle cloud at an early time, when it was still unperturbed, meaning that the exciter wire was still far from the cloud. The panel on the right shows the cloud at a later time (after the wire's motion had stopped) while a compressional pulse was propagating in the $+x$ direction.

The unperturbed microparticle cloud had the microstructure of a crystalline lattice, as can be seen in the upper-left panel of Fig.~\ref{fUnexpected2camviews}. Microparticles were self-organized into a triangular lattice with six-fold symmetry, \textit{i.e.}, it was hexagonal, with a lattice constant of about 0.46~mm. In the central portion of the cloud ($x < 25$~mm in Fig.~\ref{fUnexpected2camviews}), the number density was roughly uniform, with a variation of $\pm 10$~\%. Near the cloud's edge ($x > 30$~mm) the number density diminished to zero, as is typical for experiments with 2D microparticle clouds.

The perturbed and unperturbed conditions differed in four significant ways. First, when it was perturbed, the cloud had a microstructure that was much less crystalline, as seen in the upper-right panel of~Fig.~\ref{fUnexpected2camviews}. Second, there was some out-of-plane displacement within the perturbed cloud, as seen in the inset with $2\times$~magnification. Third, a significant compression for the perturbed condition is seen in the top-view and side-view images, and in the density profile as well. This significant compression leads us to describe the pulse as having a high amplitude. Fourth, at $x \approx 22$~mm the profile has a sharp gradient, which is one of the characteristics that are required to characterize a compressional pulse as a shock, as we will discuss below.

\subsection{Out-of-plane displacements}

When the microparticle cloud was unperturbed, it had only a single layer. This single-layer structure is seen in the middle panel of~Fig.~\ref{fUnexpected2camviews}. We never observed out-of-plane displacements in the unperturbed microparticle cloud.

However, when the cloud had been perturbed, shortly after the wire's movement ceased, we observed out-of-plane displacements in all six runs. These out-of-plane displacements generally occurred either near the peak density, or behind the peak as in~Fig.~\ref{fUnexpected2camviews}. We observed fewer out-of-plane displacements ahead of the peak.

\begin{figure}[ht]
	\includegraphics[width=0.9\columnwidth]{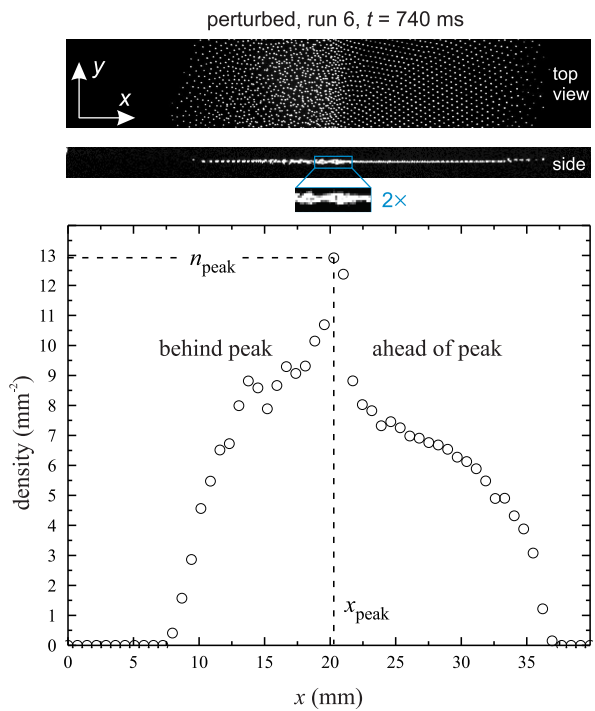}
	\caption{\label{fUnexpected3additionalStrong}Images of the microparticle cloud, and corresponding number-density profiles, in the region of interest, for run~6. This was the run that used the fastest wire speed, so that the pulse had the highest amplitude of all our runs. The data shown here are from a sequence that is presented as a video in the supplementary material.~\cite{supp}}%
\end{figure}

The width~$w_{20}$ of the region with detectable out-of-plane displacements was measured. In Table~\ref{tUnexpectedParam} we report this width,  once for each run, when the density's peak  $n_{\rm{peak}}$ was located near $x_{\rm{peak}} = 20$~mm (\textit{i.e.} $x_{\rm{peak}} \approx 20$~mm). These measurements were made manually by visually inspecting the side-view camera images (such as those in the middle panels of ~Figs.~\ref{fUnexpected2camviews}~and~\ref{fUnexpected3additionalStrong}) and assessing whether we could detect any out-of-plane displacement. The threshold of detection was essentially determined by the resolution of the side-view camera.

\begin{table*}
	\caption{\label{tUnexpectedParam}Parameters for each run. After the exciter wire was stopped, measurements of the pulse were made. The values of $n_{\rm{max}}$ correspond to the greatest  density for an entire run (\textit{i.e.}, not for a specific frame). The width of the region where out-of-plane displacements were observed is indicated as $w_{20}$, which was for the frame where the peak density was located at $x = 20$~mm, in runs~2-6.  In run~1, out-of-plane displacements were sporadic, and did not appear in that particular frame.}
	\begin{ruledtabular}
		\begin{tabular}{p{0.3in}p{0.3in}p{0.6in}p{0.5in}p{0.5in}p{0.3in}p{0.5in}p{0.6in}p{0.5in}}
			& \multicolumn{2}{c}{Exciter wire's motion}                          & \multicolumn{6}{c}{Measurements after exciter wire was stopped}                                                                                                                               \\ \cline{2-3} \cline{4-9}
			Run    & Wire speed (mm/s)            & Time wire stopped (ms)      &  $n_{\rm{max}}$~(mm${}^{-2}$)   & Compression factor $n_{\rm{max}}/n_0$ & Shock speed (mm/s)          & Mach number                & Width $w_{20}$~(mm)              & Figure                           \\ \hline
			{ 1} & { 44.5}  & { 1140} & { 10.8} & { 2.0}  & { 47.8} & { 3.0} & { -}    & { }          \\
			{ 2} & { 50.8}  & { 990}  & { 11.6} & { 2.1}  & { 52.5} & { 3.3} & { 1.37} & { \ref{fUnexpected2camviews},\ref{fUnexpected5main}} \\
			{ 3} & { 57.2}  & { 900}  & { 11.9} & { 2.2}  & { 53.6} & { 3.3} & { 0.38} & { }          \\
			{ 4} & { 63.5}  & { 800}  & { 12.4} & { 2.3}  & { 56.0} & { 3.5} & { 3.1}  & { }          \\
			{ 5} & { 76.2}  & { 700}  & { 13.3} & { 2.4}  & { 57.8} & { 3.6} & { 5.4}  & { }          \\
			{ 6} & { 101.6} & { 530}  & { 14.5} & { 2.7}  & { 66.3} & { 4.1} & { 5.4}  & { \ref{fUnexpected3additionalStrong}}   
		\end{tabular}
	\end{ruledtabular}
\end{table*}

The magnitude of the out-of-plane displacement was generally less than the horizontal distance between microparticles. Nevertheless, despite their small size, these displacements may be physically significant for explaining an energy influx mechanism, as we will discuss later.

To the best of our knowledge, ours is the first report of localized out-of-plane displacement, for a blast wave in a 2D dusty plasma. However, aside from blast waves, there have been Mach-cone shock-wave experiments where there was a prominent report of out-of-plane motion that was not only detected but also measured.~\cite{couedel2012three} We should also mention that in blast waves that were reported for previous 2D experiments, even though the authors did not mention out-of-plane displacements, we believe that we can identify some hints of this phenomenon in their reports. \citeauthor{samsonov2004shock}\cite{samsonov2004shock} noted that in their blast-wave experiment, within the horizontal sheet of laser illumination, some particles disappeared. We believe that this disappearance may have been due to out-of-plane displacements, which could have been confirmed if the experimenters had used a side-view camera like ours.

\subsection{Properties of the compressional pulse}

Two quantities that we measure, using a pulse's spatial profile, are the peak's amplitude $n_{\rm{peak}}$ and its position $x_{\rm{peak}}$. These quantities are marked in the lower-right panel of Fig.~\ref{fUnexpected2camviews}, which was for run~2, which had a slower initial velocity for the wire. For comparison, in Fig.~\ref{fUnexpected3additionalStrong}, a run with a higher initial velocity has a profile with a greater amplitude $n_{\rm{peak}}$.

\begin{figure}[ht]
	\includegraphics[width=0.9\columnwidth]{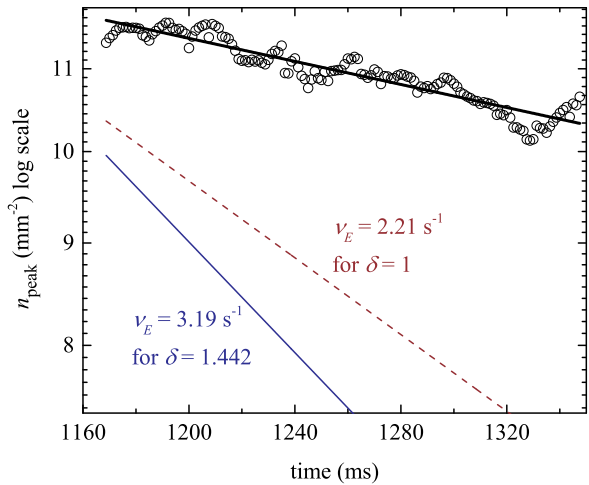}
	\caption{\label{fUnexpected5main}Peak number density $n_{\rm{peak}}$ in each video frame, obtained as in Fig.~\ref{fUnexpected2camviews}. The data shown were recorded in run~2, after the exciter wire stopped at 990~ms. The axes are semilogarithmic, so that an exponential decay would appear as a line, with a slope corresponding to a damping rate with units of~s${}^{-1}$. The experimentally observed decay, seen in the solid curve fitting the data points, has a rate of $\nu_{\rm{pulse}} = 0.66$~s${}^{-1}$. For comparison, this value is much smaller than the theoretical rate in Eq.~(\ref{eqUnexpectedNuBord}), which is bracketed by two values, shown here as representative slopes in the lower lines; these theoretical values assume that neutral gas friction alone affected the kinetic energy of microparticles.}%
\end{figure}

Measurements of the peak's amplitude $n_{\rm{peak}}$ as it developed with time are presented in Fig.~\ref{fUnexpected5main}. The data points shown are experimental results, which we obtained by selecting the greatest value of density in a given video frame. Essentially, these data points describe the blast wave's amplitude as the density measured at the wave's crest. While $n_{\rm{peak}}$ is a peak density in a specific frame, we define $n_{\rm{max}}$ as the greatest density registered in an entire run. The numerical values of  $n_{\rm{max}}$ are given in Table~\ref{tUnexpectedParam}.  For example, for run~2, $n_{\rm{max}} = 11.6$~mm${}^{-2}$.

Figure~\ref{fUnexpected5main} is our chief result; it demonstrates a lack of significant decay as the pulse propagates. During the time interval shown, the exciter wire had already stopped moving, so that the input of mechanical energy had already ceased. Despite the lack of this mechanical energy input, we see in Fig.~\ref{fUnexpected5main} that the pulse's amplitude remained nearly constant. The amplitude diminished at a slow rate,  $\nu_{\rm{pulse}} = 0.66 \pm 0.02$~s${}^{-1}$, when fit to an exponential $ \propto \exp(-\nu_{\rm{pulse}} t)$.  We will discuss this result shortly.

Measurements of the pulse's speed are presented in Table~\ref{tUnexpectedParam}. We obtained these values from measurements of the peak's location $x_{\rm{peak}}$ in each video frame, yielding a time series, $x_{\rm{peak}}$ vs time. From this time series, we obtained the pulse's propagation speed, using the method of Ref.~\cite{kananovich2020experimental}.

Table~\ref{tUnexpectedParam} shows that the pulse propagated at a supersonic speed in all six runs. This speed ranged from 48~mm/s to 66~mm/s, which was much greater than the longitudinal sound speed $c_l = 16$~mm/s. Also shown in Table~\ref{tUnexpectedParam} are values of the initial wire speed, the greatest measured density for every run $n_{\rm{max}}$, compression factor $n_{\rm{max}}/n_0$, shock speed, shock Mach number, and the width~$w_{20}$.

\section{Discussion}

\subsection{Attributes of a shock}

We can classify our pulses as shock waves because they have three key attributes of shocks: supersonic speed, sharp gradient, and high amplitude. A supersonic speed of the pulse's propagation was observed for the six experimental runs, with Mach numbers ranging from 3.0 to 4.1, as summarized in Table~\ref{tUnexpectedParam}. A sharp gradient of the propagating pulse is apparent in the lower right panel of Fig.~\ref{fUnexpected2camviews} and, most prominently, in Fig.~\ref{fUnexpected3additionalStrong}. A high amplitude of a pulse is evident in comparing the lower panels of Fig.~\ref{fUnexpected2camviews}. The high amplitude is also observed in Fig.~\ref{fUnexpected3additionalStrong}, and in the videos in the supplementary material.~\cite{supp} As a measure of the amplitude,  the compression factor $n_{\rm{max}}/n_0$ ranged from 2.0 to 2.7 over our six runs. These values for the compression factor all exceed what was previously reported in the literature for blast waves in 2D dusty plasmas. For example, $n_{\rm{max}}/n_0$  was 1.2 in the experiment of \citeauthor{samsonov2004shock}~\cite{samsonov2004shock} In that paper, the pulses were identified as shocks.

Our main result was a lack of significant decay of the shock's amplitude in Fig.~\ref{fUnexpected5main}. After the exciter wire stopped moving, mechanical work by the wire on the cloud of microparticles ceased suddenly. Thereafter, during the time interval shown in Fig.~\ref{fUnexpected5main}, the shock's amplitude remained almost constant, diminishing slowly with an experimentally measured rate $\nu_{\rm{pulse}} = 0.66$~s${}^{-1}$ when fit to an exponential.

If gas damping were the only mechanism affecting the amplitude of the blast wave, after the wire stopped moving, we would expect an exponential decay, with a predictable time constant equal to $1/{\nu_E}$. The value of $\nu_E$, for our experiment has limiting values of Eq.~(\ref{eqUnexpectedNuBord}), which are presented in Fig.~\ref{fUnexpected5main} as representative straight lines to indicate slopes, since the axes are semilogarithmic. 

A great mismatch is seen in comparing the theoretical slopes for gas-friction and the experimental dependence, in Fig.~\ref{fUnexpected5main}. The amplitude of the shock wave, as observed in the experiment, decays much slower than can be accounted for by gas drag only.

\subsection{Explaining how the shock can decay little}

We suggest that the very low level of decay of our blast wave is explained by an energy influx from ion flow. We further suggest that this energy influx could occur, for example, through the Schweigert instability.~\cite{schweigert1998plasma,piel2003waves,melzer1996structure,schweigert1996alignment}

The Schweigert instability has been experimentally confirmed~\cite{melzer1996experimental} to convert energy in the ion flow into kinetic energy of the microparticles. As we mentioned earlier, the Schweigert instability is typically observed when an ion flow passes perpendicular to a layer of microspheres, and it is most profound if the microspheres in fact do not rest strictly in a single 2D plane, but instead have noticeably large out-of-plane displacements.~\cite{melzer1996experimental,schweigert1998plasma,shveigert2000melting,melzer2014connecting} As the ions flow past a microparticle, they form a downstream wake, where there is a spatially localized concentration of positive space charge. That downstream wake can attract other microparticles, which are negatively charged. This attraction to the ion wake can lead to a particularly strong energy transfer from the ion flow to the microparticles,~\cite{melzer1996experimental,schweigert1998plasma,shveigert2000melting,melzer2014connecting} in what has been described as a nonreciprocal interaction.~\cite{schweigert1996alignment,carstensen2012charging,melzer2014connecting,melzer1996structure,schweigert1998plasma,piel2010dusty,mukhopadhyay2014experimental} Considerable kinetic energy can be added to microparticles, if they do not rest on a single plane, but instead have out-of-plane displacements.

For our experiment, it is significant that we detected out-of-plane displacements. That detection was made using our side-view camera. Images, revealing these displacements, can be seen in the right-middle panel of~Fig.~\ref{fUnexpected2camviews} and the middle panel of Fig.~\ref{fUnexpected3additionalStrong}. They can also be seen in the videos in the supplementary material.~\cite{supp} Generally, the out-of-plane displacements were observed in a region beginning where the density gradient was greatest, near the shock's density peak, and extending up to a few millimeters behind the peak. Since these displacements are located so close the peak, it seems reasonable to suggest that the Schweigert instability could add energy that helps sustain the pulse's amplitude.

\subsection{Excluding other mechanisms}

The design of our experiment largely eliminates other energy input mechanisms, besides those arising from ion flow. We can identify two such mechanisms:

The first mechanism that we avoid is gas flow, which was present in the previous 3D experiments of \citeauthor{samsonov2003kinetic}~\cite{samsonov2003kinetic} and \citeauthor{fortov2004large}~\cite{fortov2004large} In those experiments, a gas puff compressed the microparticle cloud, creating what appears to be a blast wave that did not diminish much as it propagated. For those two experiments, the gas flow's pattern and time scale for its dissipation were not measured, so that lingering effects of the gas puff cannot be excluded in explaining the lack of decay of the wave in their experiment. For our experiment, however, the gas flow was not a factor because it was very small and never varied in time.

The second mechanism that we avoid is what \citeauthor{samsonov2011tsunami} termed a ``tsunami effect.''~\cite{samsonov2011tsunami} This is an artifact of a density gradient in the undisturbed microparticle cloud. In an experiment with solitions, \citeauthor{samsonov2011tsunami} found that this effect was responsible for the absence of amplitude decay, as the solitions propagated.~\cite{samsonov2011tsunami,durniak2009steepening} The unperturbed medium had a density profile that diminished, in the direction of pulse propagation, by 2.5-fold across the microparticle cloud, in that experiment. In our experiment, however, such a gradient was not a factor. A strong gradient appeared in our density profile only at the extreme right edge of the particle cloud, at $x > 30$~mm. For the data in Fig.~\ref{fUnexpected5main}, our pulse was located always at $x < 25$~mm, avoiding this so-called tsunami effect.

\section{Conclusions}

In earlier experiments it was found that a shock wave's amplitude did not decay as much as would be expected, due to gas friction. Those experiments were performed with a dust cloud that filled a 3D volume.~\cite{fortov2004large,fortov2010complex} We have now extended this finding to a dusty plasma that has a 2D monolayer of microparticles. The amplitude of our shock was observed to diminish very little, over an extended time. If the only energy mechanism acting on the microparticle cloud were gas friction, there would be a large and predictable exponential decay, which was not observed. This experimental finding suggests a source of energy influx to the microparticle cloud, which largely cancelled the dissipative effects of gas friction.

To help determine the source of the energy influx, we take advantage of the design of the experiment, which had two advantages. Firstly, we prepared a microparticle cloud that had no strong density gradients, in the region of interest, before we manipulated the cloud with our moving wire. Secondly, and more importantly, we used a new method of blast-wave excitation, involving the movement of a thin wire, which allowed us to suddenly stop the mechanical energy input by halting the wire's motion. In this way, we avoided lingering sources of mechanical energy, as might occur in other experimental schemes relying on excitation by a gas puff.~\cite{fortov2004large,samsonov2003kinetic} The gas in our experiment was maintained at a steady condition while the exciter wire was propelled at a supersonic speed and then suddenly stopped. 

Ion flow is a source of energy in most laboratory experiments with dusty plasmas. Ion flow is capable of driving several kinds of instabilities, depending on the conditions in the dusty plasma. Because we prepared our dusty plasma as a 2D monolayer, which was perpendicular to the ion flow, it is of particular interest to consider the Schweigert instability as a mechanism to couple the energy of the ion flow to the motion of the microparticles.  This instability can be great if there is some out-of-plane displacements of the microparticles. 

To detect such displacements we used not only a top-view camera as in previous 2D shock experiments,~\cite{samsonov2004shock,kananovich2020experimental} but also a side-view camera. Although that side-view camera did not have a high spatial resolution, it was capable of registering out-of-plane displacements.  We detected such out-of-plane-displacements in all the runs reported here. The presence of these displacements lends support to the possibility that the Schweigert instability provides an energy influx, in our experiment.

\section{Acknowledgments}
The authors thank Zian Wei for technical assistance. This work was supported by U.S. Department of Energy grant DE-SG0014566, the Army Research Office under MURI Grant W911NF-18-1-0240, and NASA-JPL subcontract 1573629.

\end{document}